\documentclass[a4paper,11pt]{article}

\usepackage{pos}
\usepackage{xspace}
\usepackage{lineno}
\usepackage{hepparticles}
\usepackage{heppennames2}

\makeatletter
\xpatchcmd\@HepConStyle
  {\edef\@upcode{\updefault}}
  {\ifdefined\shapedefault\edef\@upcode{\shapedefault}\else\edef\@upcode{\updefault}\fi}
  {}{}
\makeatother

\newcommand{\CP}{\ensuremath{CP}\xspace}
\newcommand{\ctWI}{\ensuremath{c_{\PQt\PW}^{\mathrm{I}}}\xspace}
\newcommand{\ctZI}{\ensuremath{c_{\PQt\PZ}^{\mathrm{I}}}\xspace}

\newcommand{\tZq}{\ensuremath{\PQt\PZ\PQq}\xspace}
\newcommand{\ttZ}{\ensuremath{\PQt\PAQt\PZ}\xspace}
\newcommand{\ttbar}{\ensuremath{\PQt\PAQt}\xspace}
\newcommand{\ttW}{\ensuremath{\PQt\PAQt\PW}\xspace}

\newcommand{\WZ}{\ensuremath{\PW\PZ}\xspace}
\newcommand{\ZZ}{\ensuremath{\PZ\PZ}\xspace}

\newcommand{\wSM}{\ensuremath{w_{\text{SM}}}\xspace}
\newcommand{\zjets}{\ensuremath{\text{DY+jets}}\xspace}
\newcommand{\pt}{\ensuremath{p_{\text{T}}}\xspace}

\newcommand{\Singletab}{\ensuremath{\mathcal{O}_{\varphi q}^{(1)(ab)}}\xspace}
\newcommand{\Tripletab}{\ensuremath{\mathcal{O}_{\varphi q}^{(3)(ab)}}\xspace}
\newcommand{\UpOab}{\ensuremath{\mathcal{O}_{\varphi u}^{(ab)}}\xspace}
\newcommand{\DownOab}{\ensuremath{\mathcal{O}_{\varphi d}^{(ab)}}\xspace}

\newcommand{\cHqTripletLight}{\ensuremath{c_{\varphi q}^{(3)(11+22)}}\xspace}
\newcommand{\cHqTripletThird}{\ensuremath{c_{\varphi q}^{(3)(33)}}\xspace}
\newcommand{\cHqMinusLight}{\ensuremath{c_{\varphi q}^{(-)(11+22)}}\xspace}
\newcommand{\cHqMinusThird}{\ensuremath{c_{\varphi q}^{(-)(33)}}\xspace}
\newcommand{\cHuLight}{\ensuremath{c_{\varphi u}^{(11+22)}}\xspace}
\newcommand{\cHuThird}{\ensuremath{c_{\varphi u}^{(33)}}\xspace}
\newcommand{\cHdLight}{\ensuremath{c_{\varphi d}^{(11+22)}}\xspace}
\newcommand{\cHdThird}{\ensuremath{c_{\varphi d}^{(33)}}\xspace}
\newcommand{\cW}{\ensuremath{c_{W}}\xspace}
\newcommand{\cWtil}{\ensuremath{c_{\widetilde{W}}}\xspace}

\newcommand{\tWZ}{\ensuremath{\PQt\PW\PZ}\xspace}

\title{Searches for new physics breaking the \textit{CP} and flavor symmetries in the top quark section in CMS}

\author*[a]{Sergio S\'anchez Cruz}
\author{on behalf of the CMS Collaboration}
\manuallySeparateAuthors

\affiliation[a]{CERN, European Organization for Nuclear Research\\
  Geneva, Swizterland}

\emailAdd{sergio.sanchez.cruz@cern.ch}

\abstract{
  The top quark plays an important role in a number of new physics models, some of which introduce violations to some of the accidental symmetries of the SM, such as the lepton number conservation or introduce additional sources of others already broken, such as the \CP symmetry. A set of measurements is presented that probe violation of these symmetries in processes involving the top quark,
  in association with additional particles.
}

\FullConference{The European Physical Society Conference on High Energy Physics (EPS-HEP2025)\\
7-11 July 2025\\
Marseille, France\\}

\begin{document}
\maketitle

\section{Introduction}

In addition to searches for the on-shell production of particles not
present in the standard model (SM), the CMS Collaboration~\cite{CMS,CMSRun3} is putting forward a program of
indirect searches for new physics. These searches make instead use of observables that could be influenced
by the presence of particles at a higher energy scale $\Lambda$, beyond the energy reach of the experiment.
In this context, the SM effective field theory (SMEFT)~\cite{SMEFT}, an extension of the
SM that describes the effect of new particles through additional operators extending the SM Lagrangian, is a suitable
description of the experimental searches. In its most general form, the SMEFT incorporates 2499
additional dimension-6 operators, each associated with a Wilson coefficient (WC).
Since a simultaneous measurement of such a number of operators is not yet feasible, additional 
symmetries are imposed to reduce the number of operators to a manageable level. In
the top quark sector, a set of flavor assumptions has been agreed upon in the context of the LHCtopWG~\cite{LHCtopwG}.
In practice, analyses typically restrict themselves further to a subset of operators by imposing additional
assumptions, such as the conservation of the \textit{CP} symmetry and selecting operators that are relevant
to a specific topology or final state. In the top quark sector, the most comprehensive analysis so far
by CMS probes 26 independent Wilson coefficients (WCs)~\cite{TOP22006}.

However, there is no guarantee that new physics will respect those assumed symmetries. For that
reason, the CMS Collaboration is releasing a set of searches in the top quark sector that target specific
operators that violate some of these asymmetries, such as lepton or baryon number violation~\cite{BNV,LFV} or
the introduction of flavor-changing neutral currents~\cite{FCNC}. In these proceedings, we discuss the two most
recent results of this kind released by the CMS Collaboration. The first one is a  search for new physics
breaking the \textit{CP} symmetry in the associated production of top quarks with a \PZ boson, and the
second one is a search for new physics focusing on interactions involving the \PZ boson beyond the flavor symmetries
discussed in Ref.~\cite{LHCtopwG}.

\section{Analysis methodology}

Both measurements rely on the estimation of the event yields in a set of categories of interest
designed to obtain sensitivity to different new physics scenarios. The contribution to the event
yield introduced by the signals is parametrized as a function of the WCs, which allows to
perform inference on the WCs directly from the yields. Signal event yields are estimated using Monte Carlo
simulations weighted to reproduce the kinematic properties of different WC scenarios, following the approach
described in Ref.~\cite{TOP22006} and is performed using Monte Carlo simulations
with suitable weights to resemble the kinematic properties of scenarios corresponding to different WCs values.
The weights can be parametrized as

\begin{equation}\label{eq:eft_parameterization}
    w = \wSM + \sum_{i}  c_i l_i + \sum_{ij} c_i c_j q_{ij},
\end{equation}

where \wSM is the SM contribution and the sums run over the WCs. The terms
involving $l_i$ ($q_{ij}$) represent the linear (quadratic) contributions to the event weight,
which correspond to
contribution from the interference between SM and BSM diagrams (from purely BSM contributions).
We refer to the former contributions as the ``linear'' contributions and the latter as the ``quadratic''
contributions.

In addition to the signal, other processes contribute to the event yield. Events containing at least
three leptons produced in the prompt decay of a \PW or \PZ boson, or a $\PGt$ decay, are estimated
using Monte Carlo simulations. Events from other processes, such as \ttbar or \zjets,
can contribute to this selection when 
leptons coming from other sources, such as leptons produced in hadronic jets, are accepted by
the identification algorithms. This contribution is estimated by considering events passing
the same event selection but where one of the leptons fails the identification criteria, weighted
to account for the probability that a non-prompt lepton would have passed
the nominal selection criteria.

The signal is extracted by performing a likelihood fit to the observed yields. The likelihood
function is built as a product of Poisson probability mass functions with mean
equal to the expected number of events. This expectation is modified by the Wilson coefficients, which
are determined in the fit, and by a set of systematic uncertainties
that are included in the model through a set of nuisance parameters.

\section{Search for \textit{CP} violation in the production of top quarks associated with Z bosons}

\begin{figure} 
  \includegraphics[width=\textwidth]{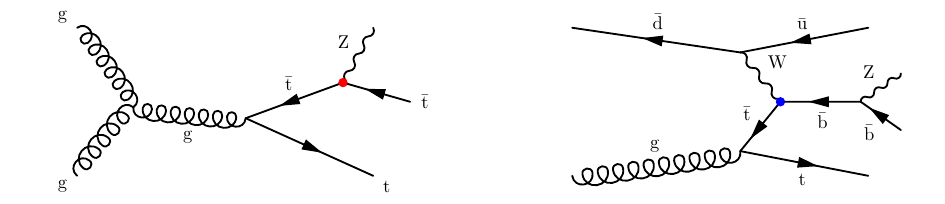}
  \caption{Representative diagrams of \ttZ (left) and \tZq production (right). Taken from~\cite{TOP24012}.}
  \label{fig:feynman_diagram}
\end{figure}

The search presented in Ref.~\cite{TOP24012} searches for \CP violation in the associated production of top quarks and \PZ bosons,
using events with three final state leptons, jets, and jets identified as a \PQb jet (\PQb-tagged jets). We use a dataset of proton-proton ($\Pp\Pp$) collisions collected
in 2016--2018 (2022), corresponding to an integrated luminosity of $138~\text{fb}^{-1}$ ($34.7~\text{fb}^{-1}$) at $\sqrt{s}=13$ TeV (13.6 TeV). Events in this selection are dominated by
the production of a top quark antiquark pair in association with a \PZ boson (\ttZ) and the production of a single top quark
in association with a \PZ boson (\tZq), which are the signal processes in this search. Some representative diagrams are shown in Fig.~\ref{fig:feynman_diagram}. Events in this selection also receive
contributions from events from $\PW\PZ$ and $\PZ\PZ$ production, and other minor top quark production processes such as \ttW or \tWZ.

\CP violation in this selection can be introduced by the \CP-odd \ctZI and \ctWI operators, defined in Ref.~\cite{TOP24012}, which
introduce additional interaction terms between the top quark and the \PZ boson,
affecting the production rate and kinematic properties of \ttZ and \tZq production, respectively. Their effect in other
processes contributing to the event yield in this selection was found to be negligible.

Due to the symmetries of \CP-odd operators, the linear contribution cancels out in any
\CP-even observable, such as the total cross section or standard kinematic variables. For that
reason, in order to be sensitive to the linear contribution, we make use of \CP-odd observables,
in which the linear contribution would manifest itself as an antisymmetric contribution to such observables.
In contrast, the SM contribution is symmetrically distributed in such observables, and so are the
most systematic uncertainties affecting its prediction. As a consequence, our search is
robust against variations in these systematic uncertainties.

\begin{figure} 
  \includegraphics[width=0.5\textwidth]{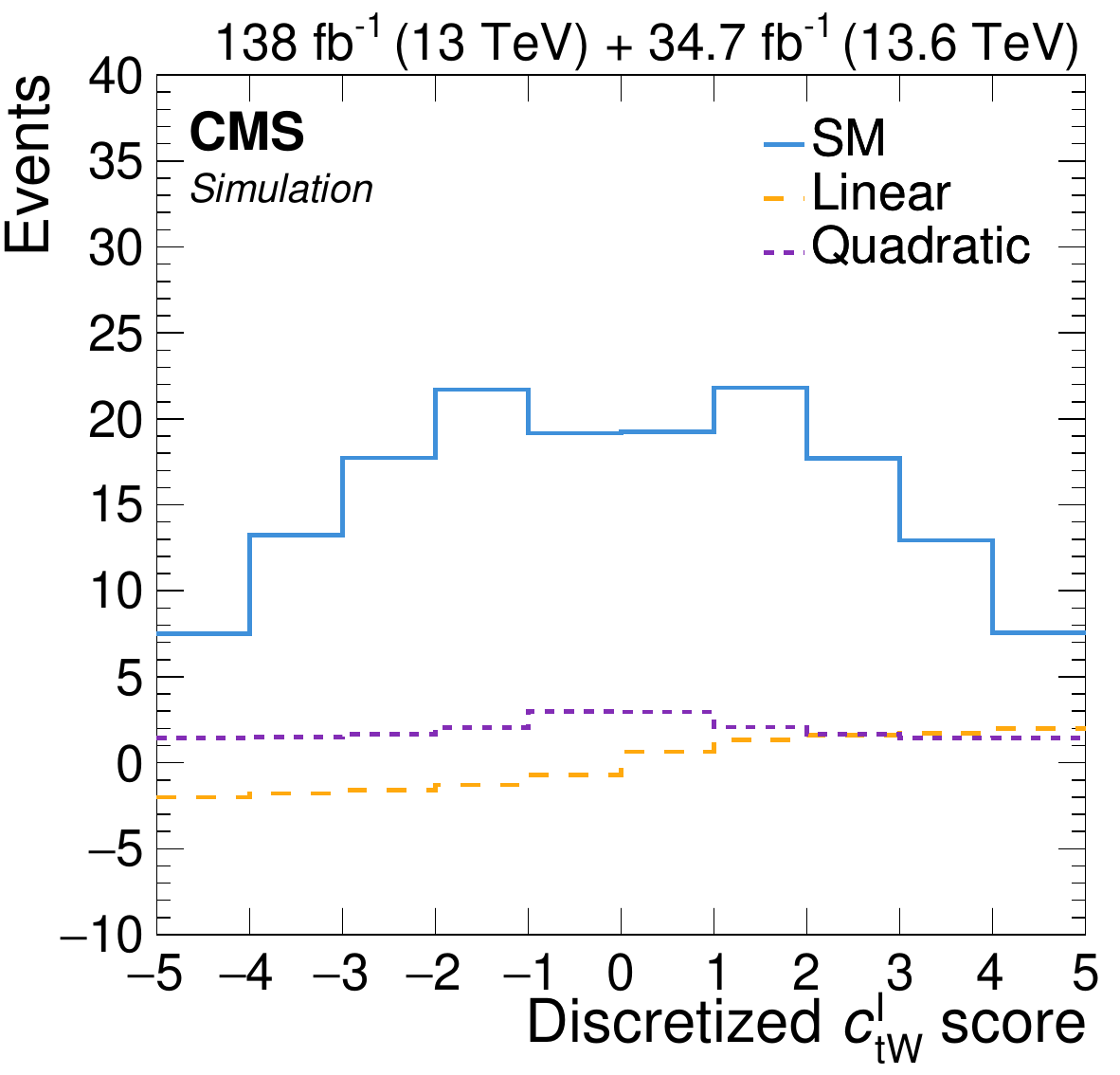}
  \includegraphics[width=0.5\textwidth]{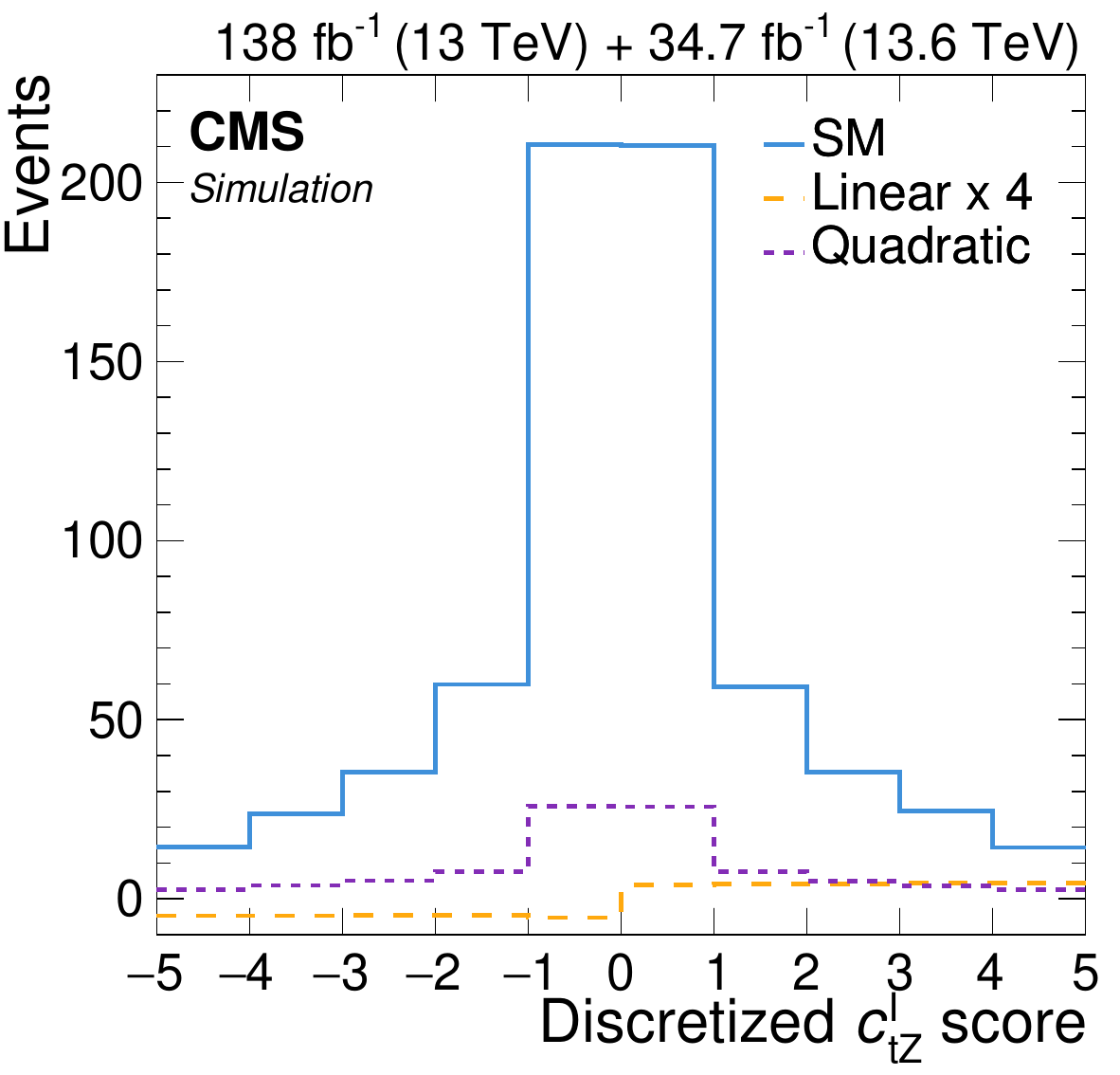}
  \caption{Distributions of $g_{\ctWI}$ (left) and $g_{\ctZI}$ (right), for \tZq and \ttZ events, respectively. Taken from~\cite{TOP24012}.}
  \label{fig:cp_distributions}
\end{figure}

In order to optimally extract sensitivity to the operators of interest, we build
two \CP-odd equivariant neural networks~\cite{eqcp}, $g_{\ctWI}$ and $g_{\ctZI}$,
one for each of the operators considered.
The network architecture 
is designed in a way that their score is odd under \CP transformations. As input variables,
we use the kinematics of the reconstructed leptons, up to 5 additional jets and missing transverse
momentum. The network is parametric as a function of the data taking era to account for the
different $\sqrt{s}$ and small changes in detector conditions. The distributions of the
neural network scores are shown in figure~\ref{fig:cp_distributions} for the signal processes
under the SM hypothesis, and the linear and quadratic terms introduced by the operators considered,
featuring the symmetry properties described above.

\begin{figure} 
  \includegraphics[width=\textwidth]{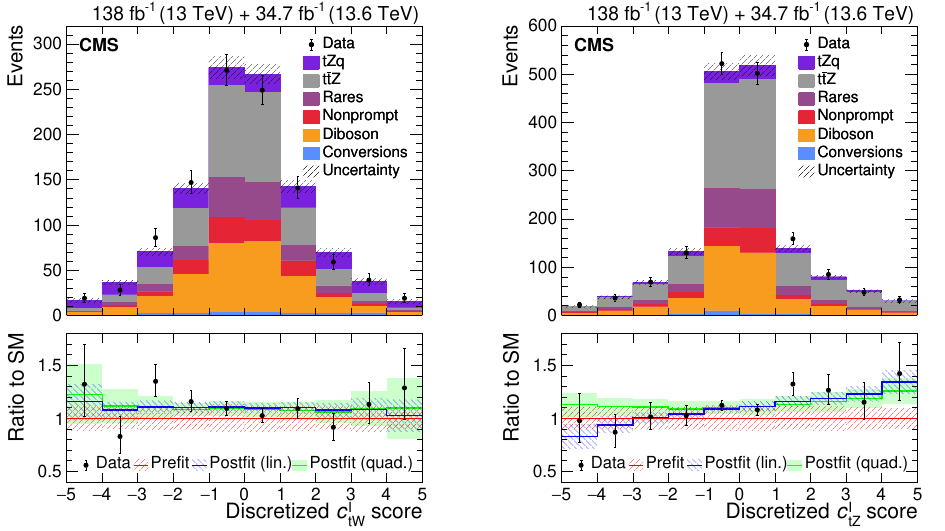}
  \caption{Distributions of $g_{\ctWI}$ (left) and $g_{\ctZI}$ (right) in the \ctWI-like and \ctZI-like regions, respectively. Taken from~\cite{TOP24012}.}
  \label{fig:cp_observations}
\end{figure}

For the signal extraction, events are split in two categories, assigning events with
$|g_{\ctWI}| > |g_{\ctZI}|$  ($|g_{\ctWI}| < |g_{\ctZI}|$ ) to the \ctWI-like (\ctZI-like) category.
Events in the \ctWI-like (\ctZI-like) category are further classified as a function of
$g_{\ctWI}$ ($g_{\ctZI}$). 
The number of observed events in bins of the two distributions are shown
in Fig.~\ref{fig:cp_observations} along with the prediction obtained by fixing all the parameters
to the value observed in the fit. Good agreement is observed between data and the prediction
for the \ctWI score in the \ctWI-like region, while a slight excess and an asymmetry are observed
in the \ctZI score in the \ctZI-like region.

\begin{figure} 
  \includegraphics[width=\textwidth]{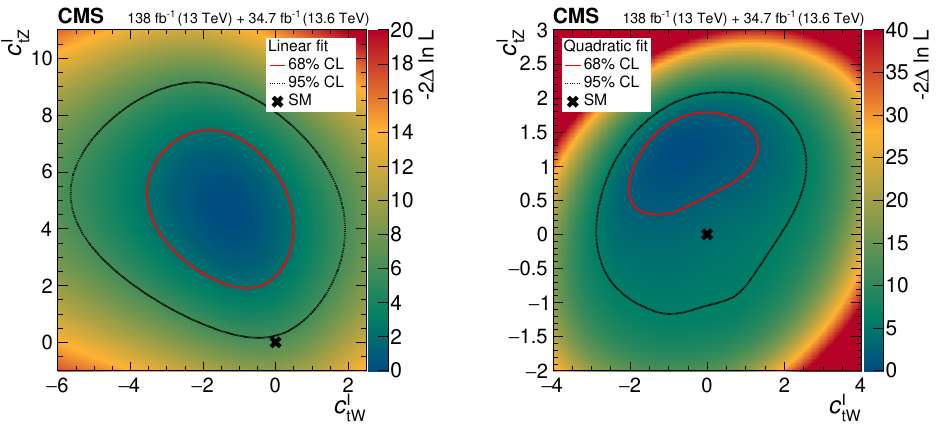}
  \caption{Likelihood scans as functions of \ctWI and \ctZI, including linear contributions only (left) and both linear and quadratic contributions (right). Taken from~\cite{TOP24012}.}
  \label{fig:cp_limits}
\end{figure}

Figure~\ref{fig:cp_limits} shows the limits as a function of \ctWI and \ctZI
for scenarios assuming the presence of linear contributions only and linear and quadratic contributions together.
The results in both scenarios are consistent with the SM within two standard deviations. 
The largest discrepancy is observed in \ctZI where data favors positive values, with an observed local significance 
of 2.5 standard deviations relative to the SM hypothesis, when only linear terms are considered. The discrepancy is driven by the
slight asymmetry observed in the $g_{\ctZI}$ distribution.

\section{Flavor structure of dimension-6 EFT operators in multilepton final states}

Reference~\cite{TOP23009} describes a complementary search for new physics in the multilepton final state, using a dataset of $\Pp\Pp$
collisions collected in 2016--2018 at $\sqrt{s}=13$ TeV, using events with at least three leptons including a
\PZ boson decay candidate. 

The analysis explores a subset of operators that
modify the interactions between \PZ bosons and quarks: $\Singletab$, $\Tripletab$, $\UpOab$, $\DownOab$, along with another
two that modify the $\PW\PZ$ vertex, namely $\mathcal{O}_W$ and $\mathcal{O}_{\widetilde{W}}$. The $a$ and $b$ indices denote
the indices of the generations affected by the operator. This analysis takes into account, for the first time, 
operators affecting different generations simultaneously. In particular, this analysis considers the associated Wilson coefficients
$\cHqTripletLight$, $\cHqTripletThird$, $\cHqMinusLight$, $\cHqMinusThird$, $\cHuLight$, $\cHuThird$, $\cHdLight$, $\cHdThird$,
$\cW$, and $\cWtil$. Operators labeled with $33$ in the superindex introduce modifications in the interactions between the top quark and the \PZ
boson, affecting the production rate and kinematic properties of \ttZ production. Operators labeled $11+22$ 
modify couplings with the first and second generation quarks, affecting the production rate and kinematics of \ttZ but also diboson processes
such as \WZ and \ZZ. The $\cW$ and $\cWtil$ coefficients modify the $\PW\PZ$ and $\PZ\PZ$ interactions and therefore affect the production rate and kinematics
of $\PW\PZ$ and $\PZ\PZ$ production. All the processes mentioned contribute to the event yield in selection used in the analysis so we perform a
simultaneous measurement of all the operators.

To gain additional sensitivity, events are categorized in three regions labeled as \ttZ, \WZ and \ZZ regions. In addition to the
requirements indicated above, events in the \ttZ region are required to have no additional leptons, at least three jets
and at least one \PQb-tagged jet. To build the \WZ region, we require instead the missing transverse momentum to be
larger than 60 GeV and no jets identified as \PQb tagged. Events in the \ZZ region are required to contain at least four leptons
forming two \PZ boson candidates. Events are further classified according to the \pt of the
leading \PZ boson candidate.

\begin{figure}[h]
  \includegraphics[width=0.3\textwidth]{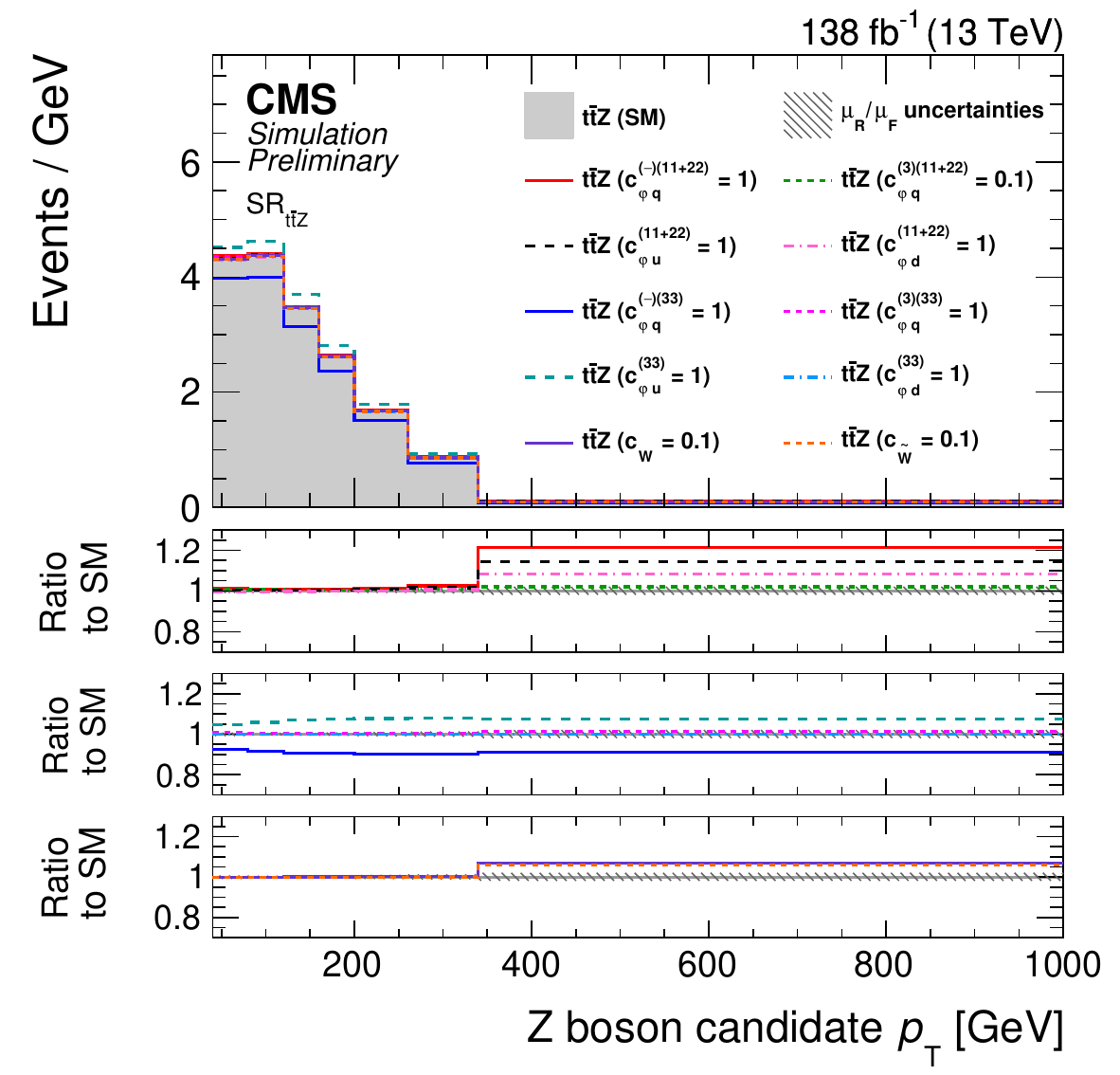}
  \includegraphics[width=0.3\textwidth]{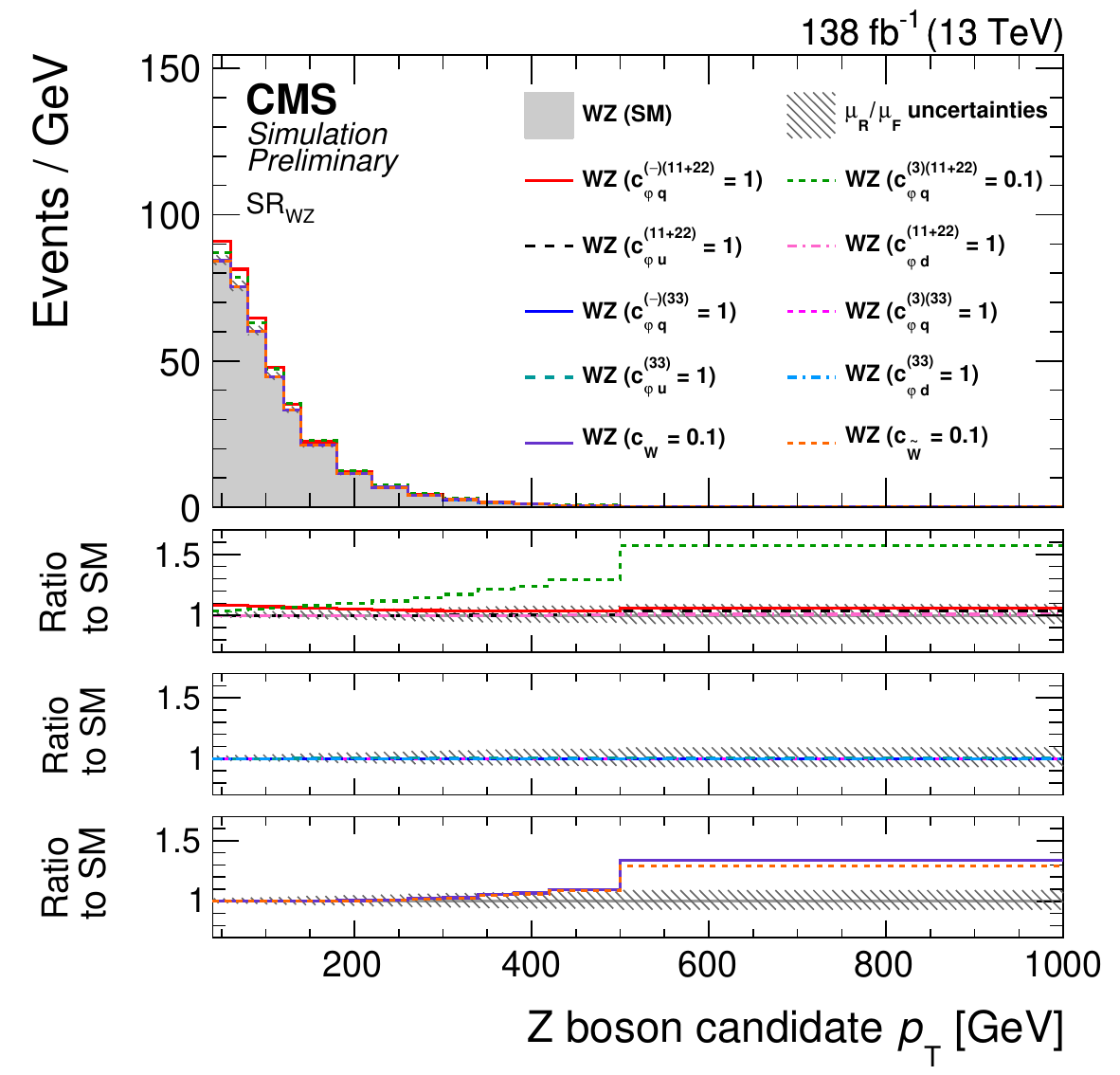} 
  \includegraphics[width=0.3\textwidth]{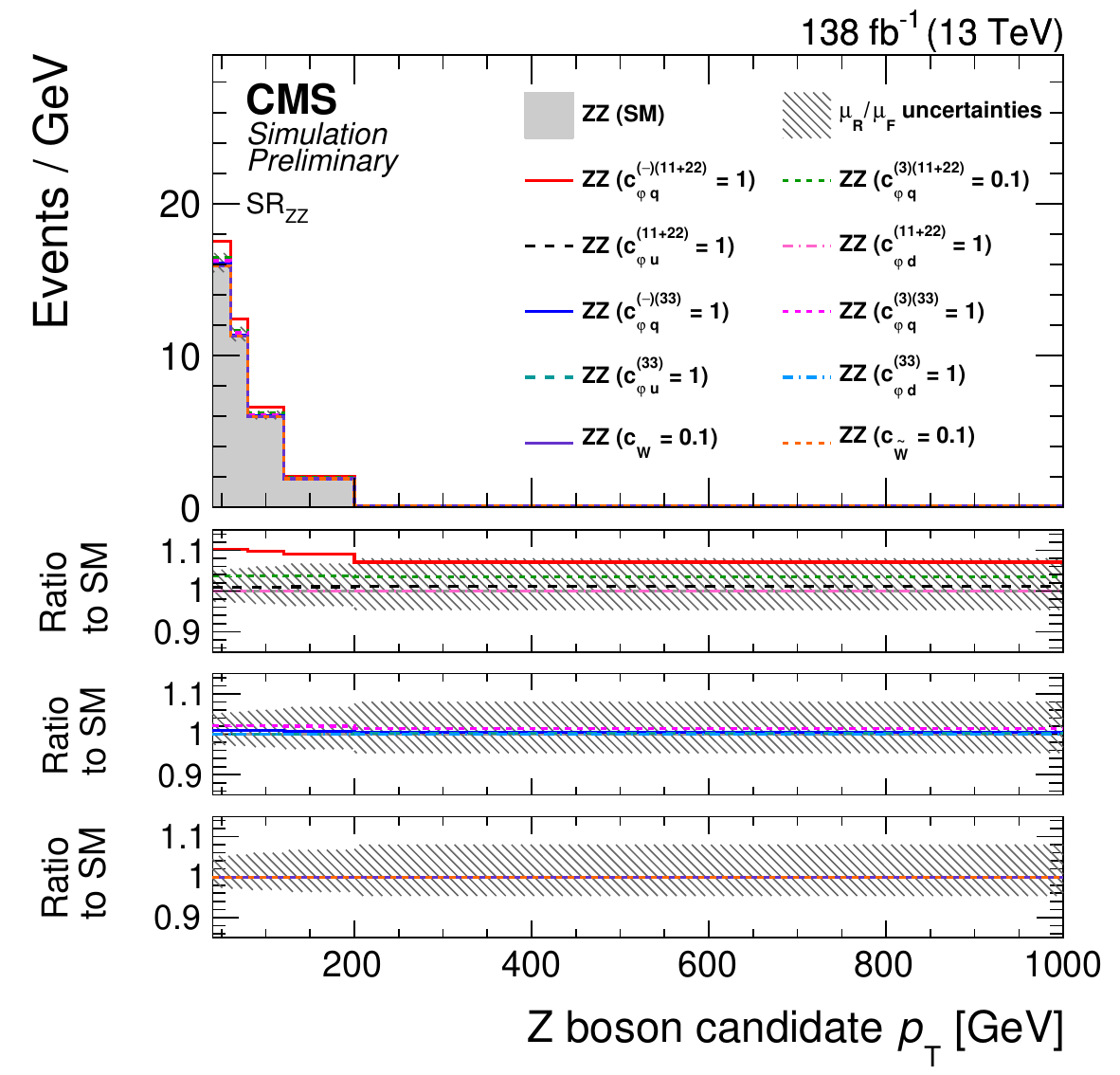}

  \includegraphics[width=0.3\textwidth]{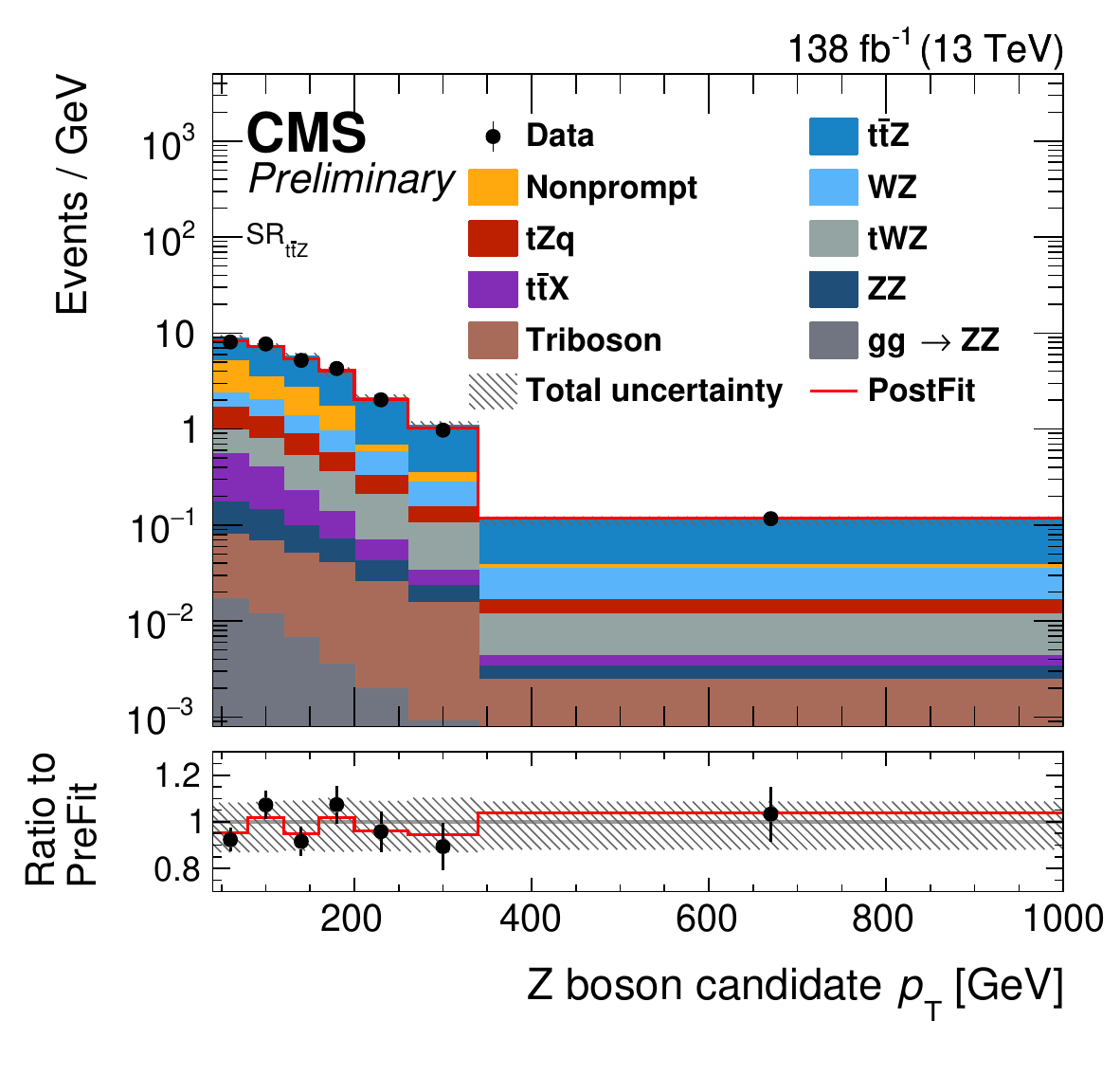}
  \includegraphics[width=0.3\textwidth]{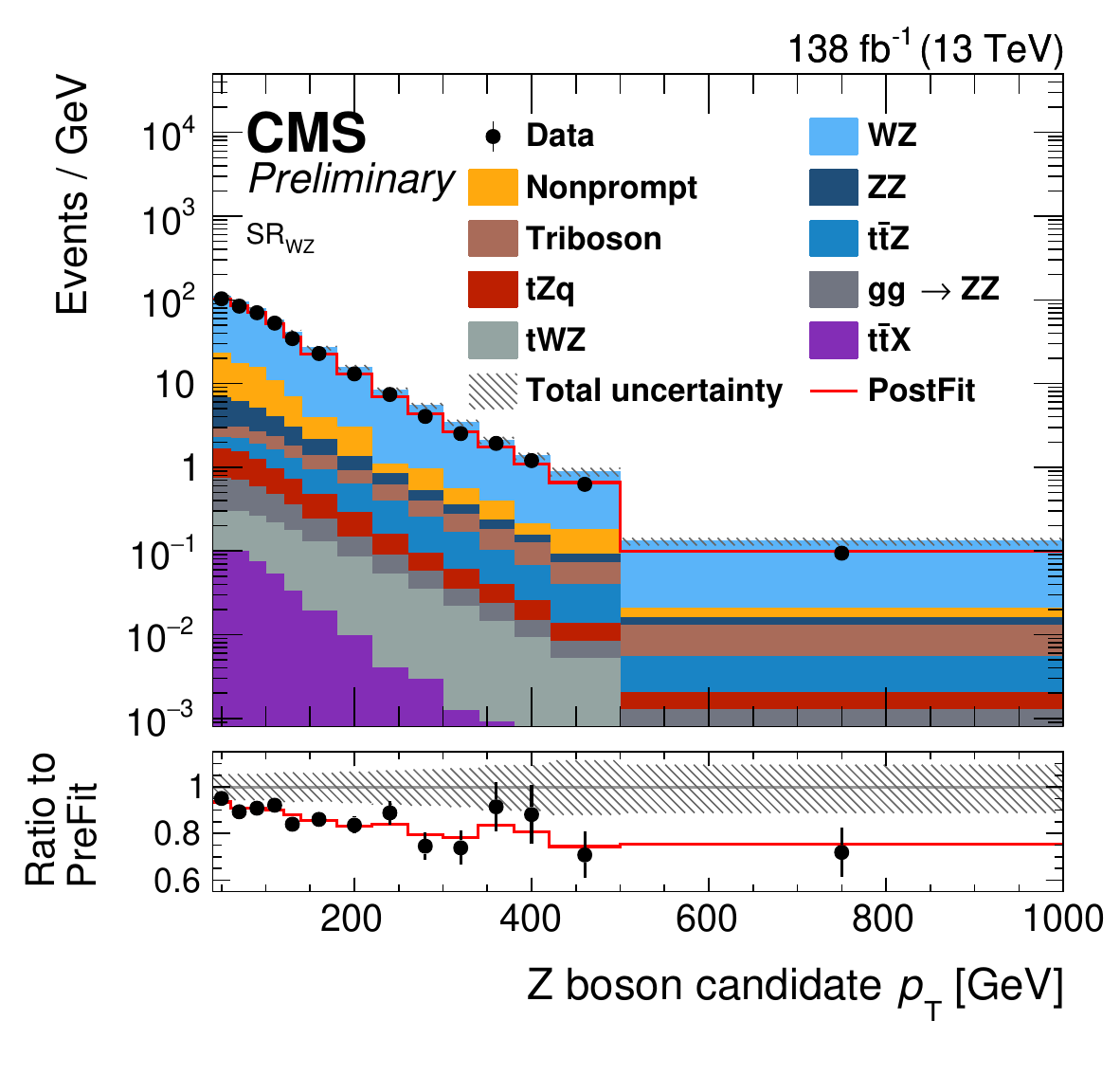} 
  \includegraphics[width=0.3\textwidth]{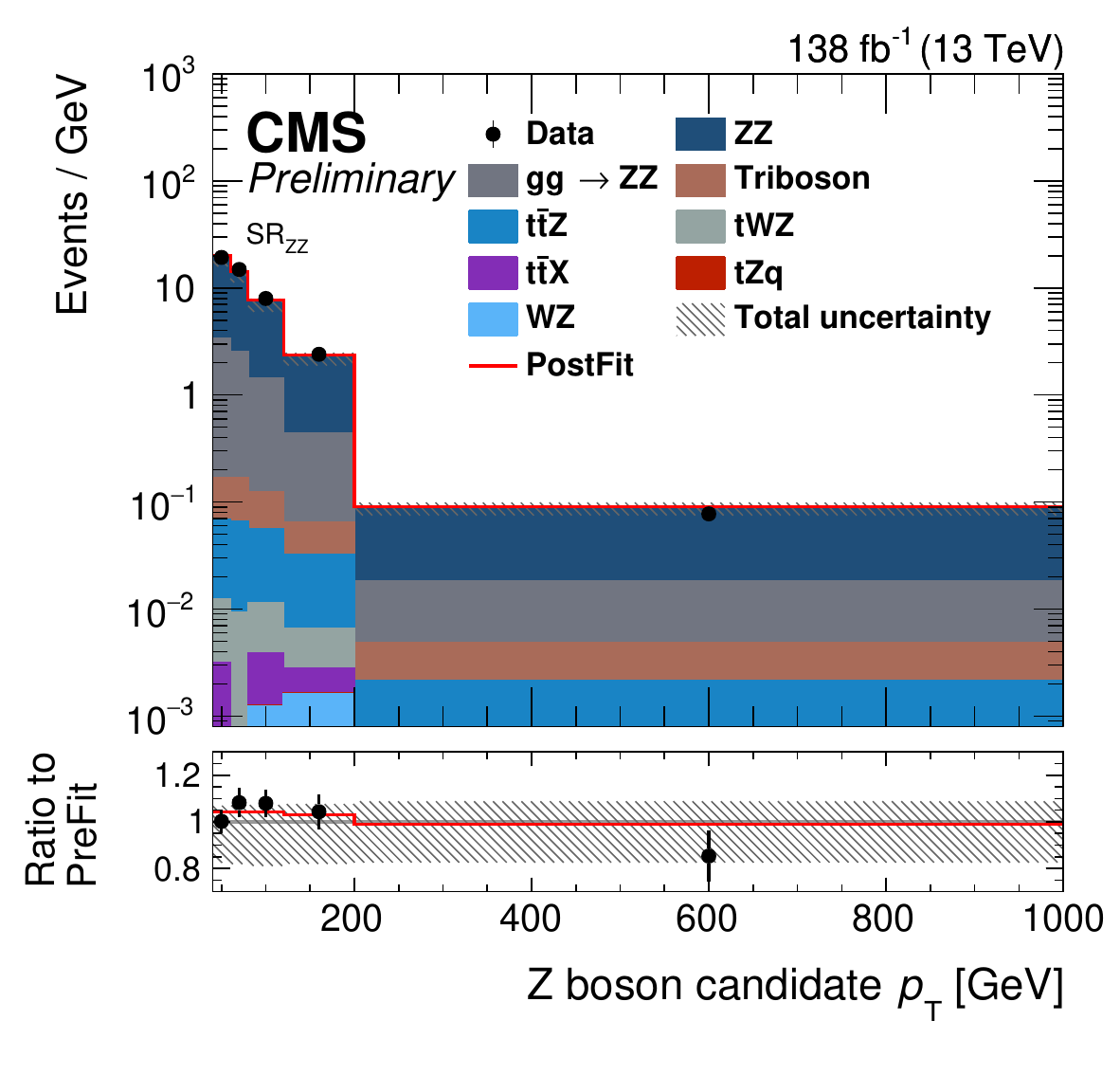} 

  \caption{Upper row: Distribution of the \PZ boson candidate \pt for \ttZ (left), \WZ (middle), and \ZZ (right) simulated events, each in its corresponding region.
    The SM prediction is shown along the prediction for different BSM scenarios. Lower row: Observed and simulated distributions of the \PZ boson candidate \pt in the
  \ttZ (left), \WZ (middle), and \ZZ (right) regions. Taken from~\cite{TOP23009}.}
  \label{fig:flavor_distributions}
\end{figure}

The effects of the operators on the production rate and \PZ boson candidate \pt distribution
are shown in Fig.~\ref{fig:flavor_distributions}. It shows that the \ttZ process is sensitive to
third generation couplings, while \WZ and \ZZ production are sensitive to the couplings of light quarks
to the \PZ bosons. The observed distributions are also shown in Fig.~\ref{fig:flavor_distributions}. Compared to the SM prediction, the \WZ
and \ZZ cross sections are measured to be 5\% lower and 10\% higher, respectively.

\begin{figure}[h!]
  \centering
  \includegraphics[width=0.485\textwidth]{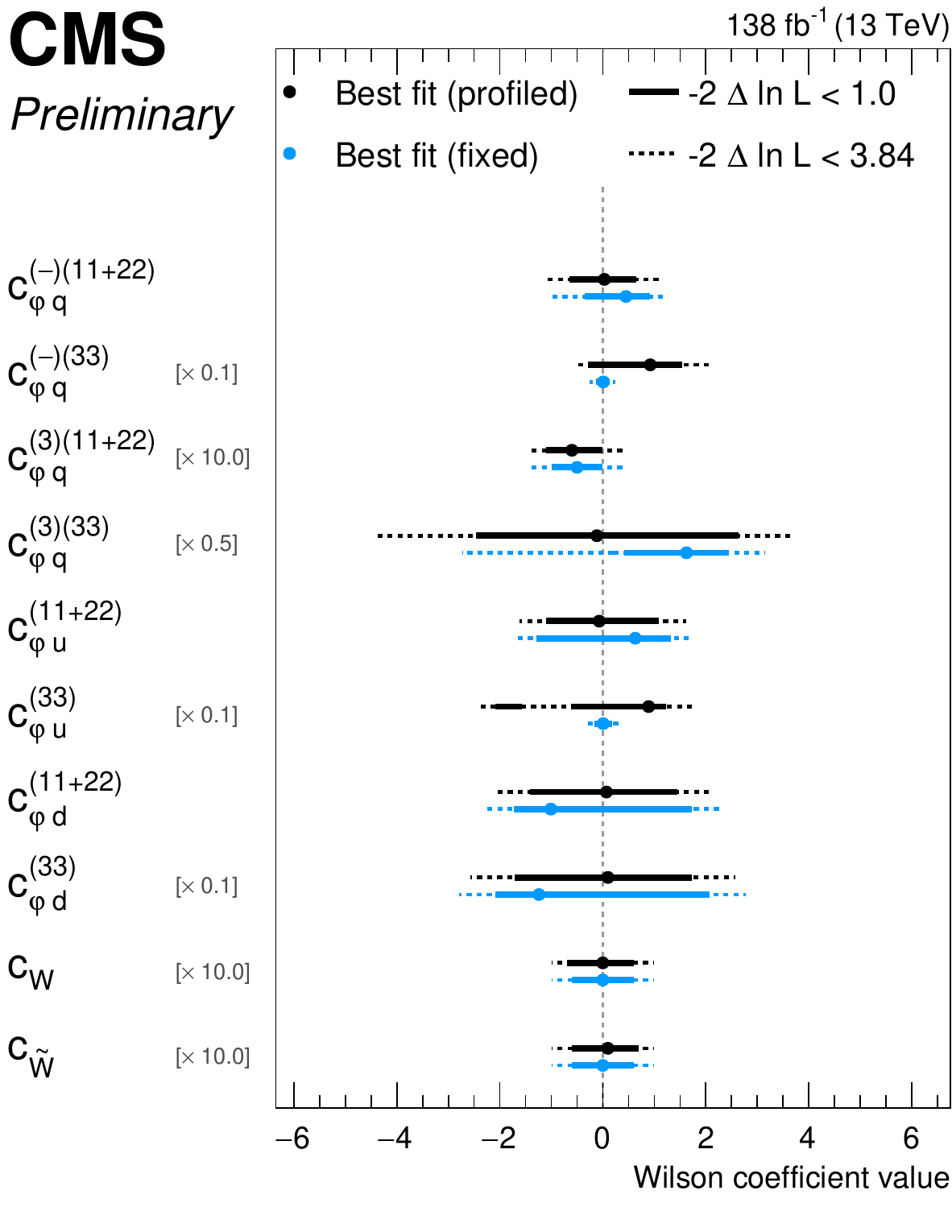}
  \caption{Observed limits on the Wilson coefficients of interest. Taken from~\cite{TOP23009}.}
  \label{fig:flavor_summary}
\end{figure}

Figure~\ref{fig:flavor_summary} summarizes the observed WC limits, both individually and in a simultaneous fit.
The slight discrepancies in the \WZ and \ZZ normalization drive small discrepancies of the $\cHqTripletThird$ and $\cHqTripletLight$
operators but are still within two standard deviations of the SM values.
Besides that, all results are consistent within one standard deviation
from the SM value.

\section{Summary}

These proceedings discussed the two most recent results from the CMS Collaboration
on indirect searches for new physics that break the \CP and flavor symmetries. No significant
deviations were reported with respect to the SM prediction. These measurements are
still dominated by the limited sizes of the LHC Run 2 and early Run 3 datasets, and are therefore
expected to become more sensitive as larger datasets are collected and analyzed.


\newpage

\begin{thebibliography}{99}
\bibitem{CMS} CMS Collaboration, \textit{JINST} \textbf{3} S08004 (2008)
\bibitem{CMSRun3} CMS Collaboration, \textit{JINST} \textbf{19}  P05064 (2024)
\bibitem{SMEFT} CMS Collaboration, \textit{Annals Phys.} 335 (2013) 21
\bibitem{LHCtopwG} J. A. Aguilar Saavedra et al., \textit{arXiv:1802.07237}
\bibitem{TOP22006} CMS Collaboration, \textit{JHEP} 12 (2023) 068
\bibitem{BNV} CMS Collaboration, \textit{PRL} 132 (2024) 241802
\bibitem{LFV} CMS Collaboration, \textit{LFV} 111 (2025) 012009
\bibitem{FCNC} CMS Collaboration, \textit{PRD} 109 (2024) 072004
\bibitem{TOP24012} CMS Collaboration, \textit{arXiv:2505.21206}. Accepted by PLB
\bibitem{eqcp} S. S\'anchez Cruz et al., \textit{PRD} 110, 096023 (2024)
\bibitem{TOP23009} CMS Collaboration, \textit{CMS-PAS-TOP-23-009}
\end{thebibliography}
\end{document}